# Detection of Surgical Site Infection Utilizing Automated Feature Generation in Clinical Notes

Feichen Shen, PhD[1], David W Larson, MD[2], James M. Naessens, ScD[1],
Elizabeth B. Habermann, PhD[1], Hongfang Liu, PhD[1], Sunghwan Sohn, PhD[1]
[1]Department of Health Sciences Research, [2]Department of Surgery
Mayo Clinic College of Medicine, Rochester MN

**Abstract**

*Postsurgical complications (PSCs) are known as a deviation from the normal postsurgical course and categorized by severity and treatment requirements. Surgical site infection (SSI) is one of major PSCs and the most common healthcare-associated infection, resulting in increased length of hospital stay and cost. In this work, we assessed an automated way to generate lexicon (i.e., keyword features) from clinical narratives using sublanguage analysis with heuristics to detect SSI and evaluated these keywords with medical experts. To further validate our approach, we also conducted decision tree algorithm on cohort using automatically generated keywords. The results show that our framework was able to identify SSI keywords from clinical narratives and to support search-based natural language processing (NLP) approaches by augmenting search queries.*

Introduction

Improving health care quality and patient safety has been a primary priority of the past 15 years since the Institute of Medicine (IOM) report [1]. Postsurgical complications (PSCs) are defined as a deviation from the normal postsurgical course and can be divided into minor and major complications or graded by severity and treatment requirements [2]. A large number of risk factors are related to PSCs including type and extent of the surgical procedure, perioperative bleeding, anesthetic method, patient's age, condition and nutritional state [3-9]. By recent statistics, there are nearly 234 million major surgical procedures are conducted all around the world each year comprises the rates of major PSCs ranging from 3% to 16% [10]. Meanwhile, rates of death or permanent disability various from 0.4% to 0.8% [11]. Therefore, detection of PSCs is a very essential step to improve patient safety and health condition as well as to reduce surgery risk and health care cost. There exist many methods for detecting postsurgical complications for patients. The Agency for Healthcare Research and Quality (AHRQ) Patient Safety Indicators (PSI) are developed based on administrative hospital ICD-9 discharge diagnoses enhanced with present on admission (POA) modifiers [12]. The American College of Surgeons National Surgical Quality Improvement Program (ACS-NSQIP) extracts clinical registry data directly from sample of medical records [13]. Consumer Reports used POA-enhanced hospital claims data to identify cases with prolonged (i.e., longer than expected) risk-adjusted post-operative lengths of stay, which served as a surrogate indicator for serious non-fatal inpatient adverse outcomes [14].

There are also many related work conducted by applying natural language processing (NLP) on electronic medical records (EMRs) to detect PSCs. Fem FizHenry et al. proposed a framework to combine structured data and natural language processing of text notes to mining information within the EMRs for detection of PSCs [15]. Harvey J.Murff et al. built a model to identify postoperative complications by analyzing EMRs with NLP. They came to a conclusion that NLP-based PSC identification achieves higher sensitivity and lower specificity compared to PSI when using a randomly selected sample of Veterans Affairs Surgical Quality Improvement Program-reviewed surgical inpatient admissions [16]. Alsara A et al. provided an automated electronic query approach to identify pertinent risk factors for postoperative acute lung injury [17]. Moreover, information extraction (IE) is widely used in NLP research. The results showed that IE is solid and promising in extracting data with various formats from unstructured reports [18-20].

In our previous work, we developed innovative semantic knowledge discovery techniques [21, 22] to perform knowledge pattern analysis for detecting associations among different complications and build a query generator clinical delivery at the point of needs. In addition, we explored sublanguage analysis [23] to find keywords for PSCs from cohort with free text reports and compared our results with a list of keywords provided by subject matter experts [24]. However, systematic evaluation of sublanguage features and actual identification of PSCs using these features have not been performed. In this paper, we expanded our previous sublanguage analysis with heuristics to refine sublanguage features (i.e., to find more appropriate keywords) relevant to surgical site infection (SSI), assessed these features based on medical expert's (an attending colorectal surgeon) judgments, and performed machine learning classification to identify patients with SSI using these features to validate our approach. Furthermore, we investigated the distribution of selected keyword concepts appeared in different postsurgical days to see if there exist any interesting patterns.

**Background**

*Colorectal Surgical Site Infections*

Colorectal surgery is performed for various diseases and often requires major reconstruction of the gastrointestinal tract [25, 26]. Because of the number of indications for the surgery and disparities in the level of involvement in colorectal resections, the range of complications differs. SSI is the most common healthcare-associated infection, accounting for 31% of all hospital inpatient infections and resulting in increased length of hospital stay and cost [27]. SSIs commonly present around the fifth postsurgical day and 5-15% of patients have such complication after colorectal surgery [28, 29].

*Sublanguage Analysis*

Sublanguage is the language of a specific domain [23]. The hypothesis behind an information extraction system is the property of inequalities of likelihood in the sublanguage. Domain Taxonomy [30] and semantic lexicon [31] are two kinds of important information in cohort. Domain Taxonomy stands for report types while semantic lexicon refers to a collection of words/phrases. Words and phrases for patients with certain PSCs can have different distributions compared to those with no complications. Detection of words/phrases with high inequality of likelihood [32] can be applied to build the base of information extraction system or augmenting the queries to identify PSCs. In this paper, we focused on the semantic lexicon and obtained inequalities of likelihood for words/phrases between cases and controls (i.e., patients with SSI vs. patients without SSI) in cohort.

**Materials and Methods**

We used part of the colorectal surgery cases performed at Mayo Clinic Rochester between 2005 and 2013. This consisted of 1,178 surgery instances with 80 SSIs annotated by a medical expert (an attending surgeon from the Department of Surgery at Mayo Clinic). This cohort has been used as the retrospective data to identify PSCs in the quality improvement project at Mayo Clinic and the keyword patterns relevant to SSIs has been also compiled by subject matter experts (Table 1)

Figure 1 gives the basic workflow of our framework. The input data are clinical narratives (i.e., clinical notes) of the patients. We first lowercased all words in clinical notes and passed these contents to MedTagger annotation tool (refer to MedTagger subsection) for the generation of annotated clinical concepts defined in UMLS [33] and Mayo Corpus. In order to alleviate noise (non-relevant concepts) and select more appropriate medical concepts we applied heuristics as follows: 1) used medical concepts in specific sections in clinical notes (i.e., Hospital Summary,

Impression/Report/Plan, Subjective, Diagnosis, Secondary Diagnoses, Problem Oriented Hospital Course), 2) used only medical concepts associated with patients (i.e., remove medical concepts associated with other than patients such as family members), 3) removed negated medical concepts. Then, we applied point-wise mutual information (PMI) [34, 35] on these concepts to automatically generate salient concepts related to SSIs and ranked them based on their inequality score (details in the subsection of Concept Keywords Mutual Information). The extracted concepts were evaluated in two ways: 1) precision at $k$ – $k$ top-ranked concepts were compared with medical expert's (attending surgeon) judgement (i.e., relevant to SSI in a degree of high, medium, low, or no) and the corresponding precision at the level of $k$ was computed, 2) we used $k$ top-ranked concepts as features in supervised machine learning (decision tree) and compared the performance using the features manually compiled from domain experts.

In addition, we analyzed selected concepts related to SSIs in the timeline of postsurgical days to examine if there exist any interesting patterns. For instance, there are 10 *wound* and 8 *cellulitis* appearred in the first day after surgery. Therefore, the concept pattern for Day 1 is *(wound (10), cellulitis (8))*.

*MedTagger*

MedTagger is an open source tool released through open health natural language processing, consisting of three major components: dictionary based concepts indexing and keyword mention lookup, pattern based information extraction, and machine learning based mention identification [36, 37]. In this study, we used MedTagger to identify medical concepts in clinical notes.

*Concept Keywords Mutual Information*

We extracted patients' clinical notes within 30 days from the surgery dates for the cohort and applied PMI to measure the association among different annotated concepts. Specifically, for the complication concept (i.e., SSI), we applied the following equation adapted from PMI to access the inequality of likelihood of concepts:

$$Inequality(c, o) = \log_2(N(c,o)) * (\log_2 \frac{N(c,o)+0.01}{N(o)} - \log_2 \frac{N(c)}{N}) \text{ (Eq 1)}$$

where $N$ is the number of surgical cases, $N(o)$ is the number of cases which contain concept $o$, $N(c, o)$ is the number of surgical cases which have complication $c$ and concept $o$, and $N(c)$ is the number of surgical cases with complication *c*. In Eq 1, we added 0.01 to smooth the formula and also penalized those concepts with low co-occurrence with the complications.

*SSI-related Concept Distribution*

We investigated the distribution of SSI-related concepts extracted by sublanguage analysis in different postsurgical days. The top *k* concepts based on PMI (Eq 1) were selected and their frequency (i.e., the number of times appeared in clinical notes within 30 days after surgery) were obtained. Then, these concepts with their frequency were aligned in different postsurgical days.

*Decision Tree Algorithm*

Decision tree algorithm [38] with stratified *10*-fold cross validation [39] was applied to detect SSIs using the concept keywords extracted by our approach. For each fold, we selected top *k* concept keywords based on PMI (Eq 1) from the training set and used them as features in machine learning. Algorithm 1 gives detailed steps of how we construct the decision tree based on these top *k* features. Specifically, we set pruning confidence as 0.25 and minimum number of instances per leaf as 2. We also compared the performance using automatically extracted concept keywords with the manually annotated keywords by the medical expert in the same framework.

---
**Algorithm 1** Decision Tree Construction (DTC)

---
Input: Training Data TD and Concept Keywords F
Output: Tree T
1. **Initialize** T = {}
2. create a node N in T
3. **IF** TD is empty
4.    terminate
5. **ELSE IF** all TD have the same post-surgical complication PC
6.    label N with PC, terminate
7. **ELSE IF** F is empty
8.    label N with the most common post-surgical complication PC (majority voting), terminate
9. Select a∈A, with the highest information gain, label N with a
10. **For** each value v of a
11.    grow a branch from N with condition a=v
12.    Let $TD_v$ be the subset of samples in TD with a=v
13.    **IF** $TD_v$ is empty
14.      attach a leaf labeled with the most common class in TD
15.    **ELSE**
16.      attach the node generated by DTC($TD_v$, A-a)
17. **return** T

**Results**

Our system was implemented by Java in Eclipse Juno Integrated Development Environment [40]. UIMA [41] based MedTagger annotator was used to extract medical concepts. Weka API was applied to run decision tree algorithm with cross validation [42].

*Top 30 SSI Concept Keywords*

Based on the measurement of PMI (Eq 1), we generated top 30 concept keywords related to SSI. Table 2 shows concepts ranked by their PMI scores as well as judgement from a medical expert (attending colorectal surgeon). As shown in Table 2, the medical expert validated that 24 out of 30 concepts were relevant to SSI and 16 out of which were highly relevant. Most non-relevant concepts are anatomical location. Top two concepts "wound infection" and "cellulitis" are matched with subject matter expert's keywords in Table1. One expert keyword, "contamination within the abdomen" was not actually appeared in our data set and so missed by our sublanguage analysis.

Table 3 shows precision at $k$ = 10, 20, and 30 based on the degree of relation to SSI (e.g., high, medium, low)—i.e., precision values considering top $k$ concepts as related to SSI, for example, precision at 10 based on a high degree, 4 out of 10 are high in medical expert judgement and so it is computed as 4/10 = 0.4. Considering top 10 concepts, 90% of them were related to SSI at least with any degree (i.e., high, medium, or low).

*Distribution of SSI-related Concepts within 30 Days after Surgery*

It is also interesting to investigate the distribution of features within a certain time period. To achieve this, we examined the distribution of SSI-related concept keywords based on top 30 concepts measured by PMI for patients with and without SSI separately in different days after surgery. Figure 2 shows distribution patterns of 16 high-related concept keywords and their frequency in postsurgical days only for patients with SSI in cohort. Out of the top 5 appeared concepts from each day, we summarized the most frequently appeared co-occurrence pairs in Table 4 for both cases. For patients with SSI, the results show that the highest co-occurrence of *wound* and *antibiotics* is 17 out of 31 days. This means that *antibiotics* and *wound* appear at the same time for 17 out of 31 days. Similarly, *wound* has a strong relationship with *dressing changes* with co-occurrence frequency 12. In addition, *wound infection* and *cellulitis* are two unique concepts that only exist as top 5 appeared concepts for patients with SSI and they obtain close relationship with *antibiotics* and *wound* respectively. For patients without SSI, the highest co-occurrence is also *wound* and *antibiotics*, which shows that those two general concepts are not good enough to distinguish SSI from non-SSI cases. In fact, antibiotics can be used to prevent infection (i.e., antibiotic prophylaxis) in colorectal surgery and does not necessarily an indication of SSI. Meanwhile, *drainage*, *incision* and *infection* occur more often in non-SSI than SSI cases and they maintain a relatively stronger relationship with *antibiotics* as well as with *wound*.

In addition, we divided 31 postsurgical days into three different partitions: 0-10 days, 11-21 days and 22-30 days. For each time period, we counted the total frequencies for each concept and chose top 5 for patients with SSI and without SSI. Table 5 and 6 give detailed statistics of this observation respectively. As shown in Table 5, for patients with SSI, *antibiotics*, *wound* and *dressing changes* are three dominant concepts in early stage. However, frequencies of these three concepts decrease from early stage to mid stage, what is more, *antibiotics* even drops out of top 5 concepts in late stage. Another interesting finding is that *wound infection* starts to increase at mid stage and decreases a lot at late stage. Furthermore, *cellulitis* increases at the late stage. In general, both *wound infection* and *cellulitis* appear in this case and they are defined as the keywords by the expert in Table 1.

For the case of patients without SSI, *antibiotics*, *incision*, *infection* and *drainage* are four concepts appear in each period. However, none of them come from the keywords given by the expert in Table 1. What is more, *incision* is denoted as a non-SSI-related concept as showed in Table 2.

*Classification on Cohort with Different Features*

Among 1,178 patients, ratio between positive (i.e., SSI) and negative (i.e., no SSI) instances is 80:1098. Top 10 features measured by PMI in Table 2 were chosen to train and test the decision tree. Moreover, we performed training and testing for 1-10 feature selections based on top 1-10 PMI rankings and computed precision, recall and F1 score for each case. To compare our sublanguage features with subject matter expert's recommendation (Table 1), we also did a decision tree classification on three features in Table 1.

Table 7 shows the performance of decision tree algorithm with different features. We observed that F1 score was highest when 4 features were applied. We also noticed that F1 score generated by 2, 3 and 5 features did not hold significant differences with which generated by experts' features.

**Discussion**

In this study, we investigated automated generation of lexicon (concept keywords) using mutual information combined with heuristics to detect SSIs. Most extracted concept keywords were relevant to SSIs based on expert review with precision of 0.90. Our approach was able to generate top ranked keywords matched with those of subject matter experts and also capture additional informative keywords that were missed by subject matter experts.

We investigated the distributions of SSI-related concepts up to 30 days after surgery. The general concepts (e.g., antibiotics, infection, wound) do not seem to be discriminative enough to differentiate SSI from non-SSI cases. However, there exist particular concepts more related to SSI (e.g., wound infection, cellulitis) than non-SSI cases. Further analysis about detailed descriptions associated with general concepts would be helpful to better identify SSI cases. These results showed the potential value to be used as a base to develop information extraction system by applying discovered distributions to augment search queries.

We applied decision tree on the colorectal surgery cohort to identify patients with SSI and compared the precision, recall and F-measure with different sets of concept keywords as well as compared the results with experts' keyword patterns. The results showed that our approach was able to produce the similar performance with the same or lower number of keywords compared to expert-based keywords (2 and 3 features in Table 7), and outperformed the performance of experts' keywords when using appropriately augmented sets of keywords (4 features in Table 7).

One limitation of this study is that we only used clinical notes to detect SSIs. The SSI patients in our cohort were actually annotated by reviewing various types of EMRs including radiology reports, operation notes, and lab. This might cause relatively low F1 score in SSI detection. However, we were able to demonstrate the capability of our approach to automatically generate lexicon that can be used as salient features to identify SSIs and promote developing advanced informatics tools for postsurgical complication detection.

The number of positive instances (surgery cases with SSI, n=80) in our cohort is relatively small for a machine learning application to fully take advantage of its potential.. In a real-world setting, it is challenging to obtain a large number of SSI cases due to the limited resource of manual annotation and low incidences of SSI cases. However, we believe that our results demonstrate the potential of automated lexicon generation to identify PSC cases and its validity. In the future, more patient data will be included and cross-institutional datasets will be used to evaluate the generalizability of the proposed approach.

We will conduct further research on the use of machine learning approaches as well as sublanguage-supported techniques to obtain higher performance of postsurgical complication detection. Moreover, we plan to expand our work to other postsurgical complications using not only clinical notes but also radiology reports, operation notes, and lab data [43].

**Acknowledgement**

This work was made possible by joint funding from National Institute of Health grants R01GM102282A1 and NIH R01 EB19403.

**References**

[1] Human TEI. Building a safer health system. Institute of Medicine. 2000;112.
[2] Dindo D, Demartines N, Clavien P-A. Classification of surgical complications: a new proposal with evaluation in a cohort of 6336 patients and results of a survey. Annals of surgery. 2004;240:205.
[3] Gombotz H, Knotzer H. Preoperative identification of patients with increased risk for perioperative bleeding. Current Opinion in Anesthesiology. 2013;26:82-90.
[4] Nielsen AW, Helm MC, Kindel T, Higgins R, Lak K, Helmen ZM, et al. Perioperative bleeding and blood transfusion are major risk factors for venous thromboembolism following bariatric surgery. Surgical endoscopy. 2017:1-8.
[5] Pereira HO, Rezende EM, Couto BRGM. Length of preoperative hospital stay: a risk factor for reducing surgical infection in femoral fracture cases. Revista Brasileira de Ortopedia (English Edition). 2015;50:638-46.
[6] Seymour DG, Pringle R. Post-operative complications in the elderly surgical patient. Gerontology. 1983;29:262-70.
[7] Garth A, Newsome C, Simmance N, Crowe T. Nutritional status, nutrition practices and post‐operative complications in patients with gastrointestinal cancer. Journal of human nutrition and dietetics. 2010;23:393-401.
[8] Faciszewski T, Winter RB, Lonstein JE, Denis F, Johnson L. The surgical and medical perioperative complications of anterior spinal fusion surgery in the thoracic and lumbar spine in adults: a review of 1223 procedures. Spine. 1995;20:1592-9.
[9] Tuman KJ, McCarthy RJ, March RJ, DeLaria GA, Patel RV, Ivankovich AD. Effects of epidural anesthesia and analgesia on coagulation and outcome after major vascular surgery. Anesthesia & Analgesia. 1991;73:696-704.
[10] Weiser TG, Haynes AB, Molina G, Lipsitz SR, Esquivel MM, Uribe-Leitz T, et al. Size and distribution of the global volume of surgery in 2012. World Health Organization Bulletin of the World Health Organization. 2016;94:201.
[11] Organization WH. New checklist to help make surgery safer. Press release: New checklist to help make surgery safer Geneva: World Health Organization. 2008.
[12] Romano PS, Mull HJ, Rivard PE, Zhao S, Henderson WG, Loveland S, et al. Validity of selected AHRQ patient safety indicators based on VA National Surgical Quality Improvement Program data. Health services research. 2009;44:182-204.
[13] Birkmeyer JD, Shahian DM, Dimick JB, Finlayson SR, Flum DR, Ko CY, et al. Blueprint for a new American College of Surgeons: national surgical quality improvement program. Journal of the American College of Surgeons. 2008;207:777-82.
[14] Fry DE, Pine M, Jones BL, Meimban RJ. Adverse outcomes in surgery: redefinition of postoperative complications. The American Journal of Surgery. 2009;197:479-84.
[15] FitzHenry F, Murff HJ, Matheny ME, Gentry N, Fielstein EM, Brown SH, et al. Exploring the frontier of electronic health record surveillance: the case of post-operative complications. Medical care. 2013;51:509.
[16] Murff HJ, FitzHenry F, Matheny ME, Gentry N, Kotter KL, Crimin K, et al. Automated identification of postoperative complications within an electronic medical record using natural language processing. Jama. 2011;306:848-55.

[17] Singh B, Singh A, Ahmed A, Wilson GA, Pickering BW, Herasevich V, et al. Derivation and validation of automated electronic search strategies to extract Charlson comorbidities from electronic medical records. Mayo Clinic Proceedings: Elsevier; 2012. p. 817-24.
[18] Friedman C. A broad-coverage natural language processing system. Proceedings of the AMIA Symposium: American Medical Informatics Association; 2000. p. 270.
[19] Sager N, Friedman C, Lyman MS. Medical language processing: computer management of narrative data. 1987.
[20] Wang Y, Wang L, Rastegar-Mojarad M, Moon S, Shen F, Afzal N, et al. Clinical Information Extraction Applications: A Literature Review. Journal of biomedical informatics. 2017.
[21] Shen F, Liu H, Sohn S, Larson DW, Lee Y. BmQGen: Biomedical query generator for knowledge discovery. Bioinformatics and Biomedicine (BIBM), 2015 IEEE International Conference on: IEEE; 2015. p. 1092-7.
[22] Shen F, Liu H, Sohn S, Larson DW, Lee Y. Predicate Oriented Pattern Analysis for Biomedical Knowledge Discovery. Intelligent Information Management. 2016;8:66-85.
[23] Kittredge R, Lehrberger J. Sublanguage: Studies of language in restricted semantic domains: Walter de Gruyter; 1982.
[24] Liu H, Sohn S, Murphy S, Lovely J, Burton M, Naessens J, et al. Facilitating post-surgical complication detection through sublanguage analysis. AMIA Summits on Translational Science Proceedings. 2014;2014:77.
[25] Hintze R, Adler A, Veltzke W, Abou-Rebyeh H. Endoscopic access to the papilla of Vater for endoscopic retrograde cholangiopancreatography in patients with billroth II or Roux-en-Y gastrojejunostomy. Endoscopy. 1997;29:69-73.
[26] Cheng Y-S, Li M-H, Chen W-X, Chen N-W, Zhuang Q-X, Shang K-Z. Complications of stent placement for benign stricture of gastrointestinal tract. World journal of gastroenterology. 2004;10:284.
[27] Control CfD, Prevention. Surgical site infection (SSI) event. Procedure Associated Module (SSI) Centers for Disease Control and Prevention (CDC), Atlanta, GA. 2015:1-26.
[28] Complications. Found at: http://www.fascrs.org/physicians/education/core_subjects/2011/Complications. Accessed Jan 2018.
[29] Society of Interventional Radiology. Found at: http://www.sirweb.org/. Accessed Jan 2018.
[30] Taylor A. Extracting knowledge from biological descriptions. Proceedings of 2nd International Conference on Building and Sharing Very Large-Scale Knowledge Bases1995. p. 114-9.
[31] Johnson SB. A semantic lexicon for medical language processing. Journal of the American Medical Informatics Association. 1999;6:205-18.
[32] Hardy G, Littlewood J, Pólya G. Inequalities. Reprint of the 1952 edition. Cambridge Mathematical Library. Cambridge University Press, Cambridge; 1988.
[33] Bodenreider O. The unified medical language system (UMLS): integrating biomedical terminology. Nucleic acids research. 2004;32:D267-D70.
[34] Church KW, Hanks P. Word association norms, mutual information, and lexicography. Computational linguistics. 1990;16:22-9.
[35] Cover TM, Thomas JA. Elements of information theory: John Wiley & Sons; 2012.
[36] Torii M, Wagholikar K, Liu H. Using machine learning for concept extraction on clinical documents from multiple data sources. Journal of the American Medical Informatics Association. 2011;18:580-7.
[37] Liu H, Bielinski SJ, Sohn S, Murphy S, Wagholikar KB, Jonnalagadda SR, et al. An information extraction framework for cohort identification using electronic health records. AMIA Summits on Translational Science Proceedings. 2013;2013:149.
[38] Friedl MA, Brodley CE. Decision tree classification of land cover from remotely sensed data. Remote sensing of environment. 1997;61:399-409.
[39] Kohavi R. A study of cross-validation and bootstrap for accuracy estimation and model selection. Ijcai: Stanford, CA; 1995. p. 1137-45.
[40] Eclipse Juno Integrated Development Environment. Found at: https://www.eclipse.org/juno/. Accessed by Jan 2018.
[41] Apache UIMA. Found at: https://uima.apache.org/.
[42] Weka. Found at: http://www.cs.waikato.ac.nz/ml/weka/.
[43] Mangram AJ, Horan TC, Pearson ML, Silver LC, Jarvis WR, Committee HICPA. Guideline for prevention of surgical site infection, 1999. American journal of infection control. 1999;27:97-134.

Table 1. Keyword patterns (as regular expressions) identified by subject matter experts. "\W" means punctuations, "\w+" means one or more letters, "|" means the choice between expression before and after the operator, "\s+" means one or more blank spaces, and P? means the pattern P occurs zero or once.

| Postsurgical Complication | Keyword patterns |
|---|---|
| surgical site infection | wound(\W|\s+)?infection; cellulitis; contamination within the abdomen |

Table 2. Top 30 concept keywords automatically extracted by sublanguage analysis. In Judgement from Expert, h, m, l and n stand for high, medium, low and no relationship with SSI, respectively

| Rank | Concepts | PMI | Judgement from Expert |
|---|---|---|---|
| 1 | wound infection | 19.37 | h |
| 2 | cellulitis | 18.62 | h |
| 3 | dressing changes | 17.03 | m |
| 4 | packed | 16.57 | h |
| 5 | wound | 14.96 | l |
| 6 | dressings | 14.48 | l |
| 7 | superficial wound | 14.45 | h |
| 8 | erythema | 14.29 | m |
| 9 | wet | 14.25 | l |
| 10 | midline | 14.03 | n |
| 11 | antibiotics | 13.72 | h |
| 12 | abdominal wound | 13.64 | n |
| 13 | midline incision | 13.60 | n |
| 14 | keflex | 12.82 | h |
| 15 | dressing change | 12.80 | m |
| 16 | infection | 12.72 | h |
| 17 | wound care | 11.60 | l |
| 18 | drainage | 11.31 | m |
| 19 | packing | 11.24 | h |
| 20 | cultures | 11.05 | h |
| 21 | incision | 10.96 | n |
| 22 | levofloxacin | 10.89 | h |
| 23 | creation | 10.81 | n |
| 24 | fluconazole | 10.70 | h |
| 25 | vac | 10.69 | h |
| 26 | pus | 10.69 | h |
| 27 | antibiotic | 10.67 | h |
| 28 | augmentin | 10.62 | h |
| 29 | surgical site infection | 10.62 | h |
| 30 | nutrition | 10.57 | n |

Table 3. Precision at10, 20, and 30 for SSI-related concept keywords

| Precision at | High relation | High or medium relation | Any relation |
|---|---|---|---|
| 10 | 0.40 | 0.60 | 0.90 |
| 20 | 0.45 | 0.65 | 0.85 |
| 30 | 0.53 | 0.67 | 0.80 |

Table 4. Co-occurrence pairs frequently occurred within 30 days after surgery. The left two columns show top co-occurrence concept pairs with their co-occurrence frequency for patients with SSI and the right two columns show which for patients without SSI. The number in parenthesis denotes how many days the concept appeared within 31 days after surgery.

| Concept Pairs for Patients with SSI | Co-occurrence frequency (appeared times/month) for Patients with SSI | Concept Pairs for Patients without SSI | Co-occurrence frequency (appeared times/month) for Patients without SSI |
|---|---|---|---|
| wound (26), antibiotics (20) | 17 | antibiotics (30), wound (24) | 22 |
| wound (26), dressing changes (17) | 12 | antibiotics (30), drainage (21) | 21 |
| wound (26), wound infection (13) | 9 | wound (24), drainage (21) | 15 |
| antibiotics (20), wound infection (13) | 8 | antibiotics (30), incision (14) | 12 |
| wound infection (13), dressing changes (17) | 7 | wound (24), incision (14) | 11 |
| wound (26), cellulitis (6) | 6 | antibiotics (30), infection (11) | 11 |

Table 5. Top 5 concepts frequencies for 3 periods within 31 postsurgical days (Patients with SSI)

| 0-10 days Top Concepts | Frequency | 11-21 days Top Concepts | Frequency | 22-30 days Top Concepts | Frequency |
|---|---|---|---|---|---|
| antibiotics | 79 | antibiotics | 43 | wound | 36 |
| wound | 78 | wound | 32 | drainage | 24 |
| dressing changes | 40 | wound infection | 30 | abdominal wound | 15 |
| dressing | 29 | dressing changes | 30 | wound infection | 12 |
| infection | 29 | drainage | 25 | cellulitis | 12 |

Table 6. Top 5 concepts frequencies for 3 periods within 31 postsurgical days (Patients without SSI)

| 0-10 days Top Concepts | Frequency | 11-21 days Top Concepts | Frequency | 22-30 days Top Concepts | Frequency |
|---|---|---|---|---|---|
| antibiotics | 123 | antibiotics | 92 | antibiotics | 47 |
| incision | 117 | infection | 47 | drainage | 47 |
| wound | 95 | drainage | 47 | wound | 35 |
| drainage | 61 | cultures | 47 | incision | 26 |
| infection | 50 | incision | 46 | infection | 21 |

Table 7. Performance for decision tree algorithm with different features

| Features used | Precision | Recall | F1 Score |
|---|---|---|---|
| Top 1 feature | 0.73 | 0.46 | 0.56 |
| Top 2 features | 0.67 | 0.73 | **0.70** |
| Top 3 features | 0.69 | 0.74 | **0.71** |
| Top 4 features | 0.81 | 0.74 | **0.77** |
| Top 5 features | 0.65 | 0.75 | **0.70** |
| Top 6 features | 0.60 | 0.60 | 0.60 |
| Top 7 features | 0.63 | 0.64 | 0.63 |
| Top 8 features | 0.60 | 0.75 | 0.67 |
| Top 9 features | 0.65 | 0.61 | 0.63 |
| Top 10 features | 0.65 | 0.65 | 0.65 |
| Experts' features | 0.67 | 0.73 | 0.70 |

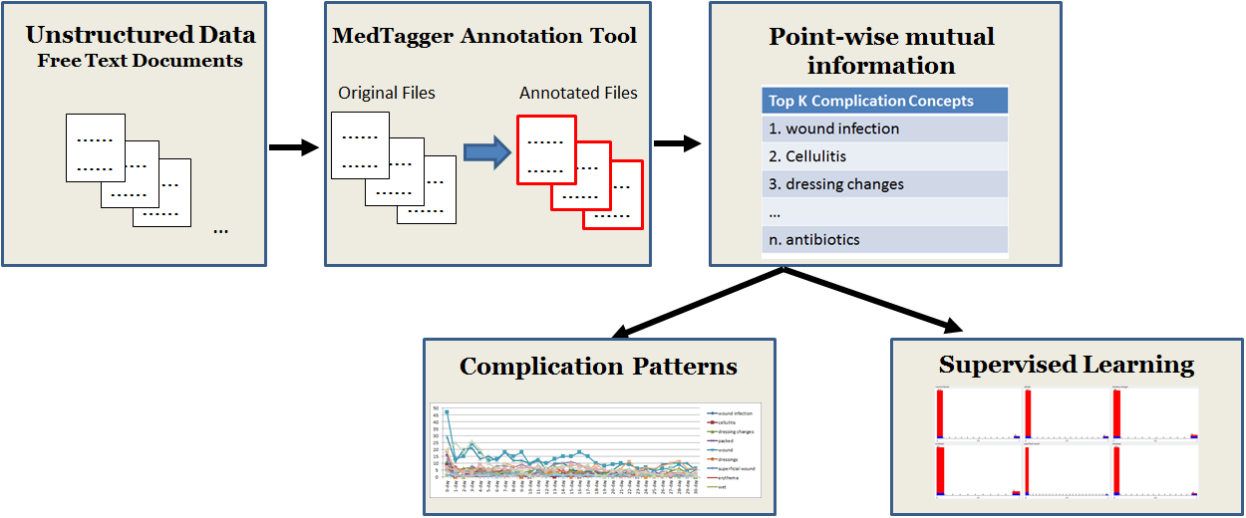

Figure 1. System workflow of surgical site infection detection

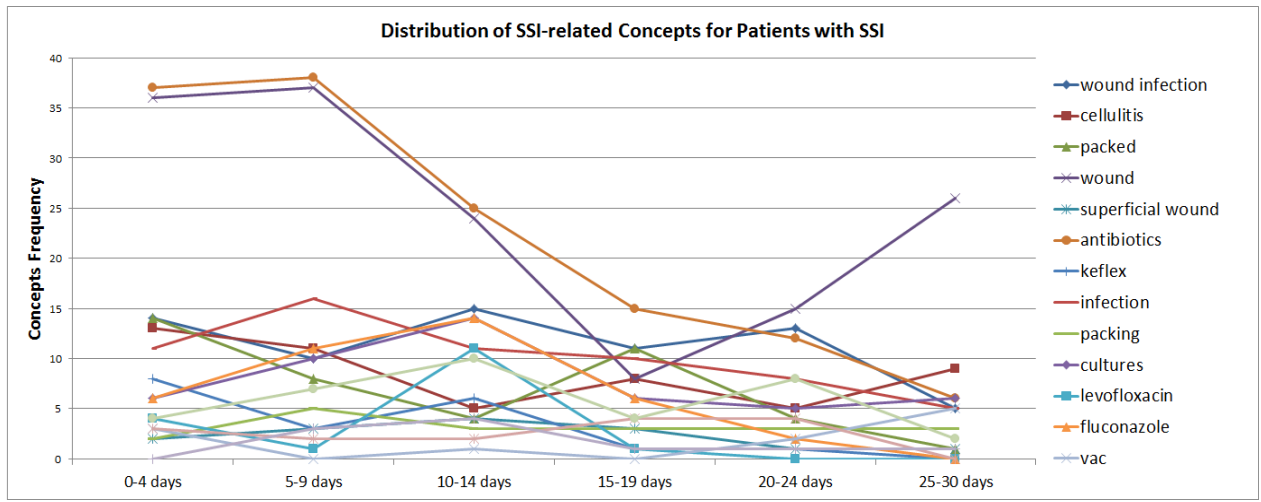

Figure 2. The distribution of SSI-related concepts in clinical notes after surgery